\documentclass[12pt]{iopart}
\usepackage{graphicx}

\begin{document}

\title[New Journal of Physics]{Dynamical structure factors of a two-dimensional Fermi superfluid within random phase approximation}

\author{Huaisong Zhao, Xiaoxu Gao, Wen Liang, Peng Zou \footnote[1] \& Feng Yuan \footnote[2]{Author to whom any correspondence should be addressed.}}
\address{College of Physics, Qingdao University, Qingdao 266071, People's Republic of China}
\ead{phy.zoupeng@gmail.com; yuanfengqdu@163.com}
\vspace{10pt}
\begin{indented}
\item[]May 2020
\end{indented}

\begin{abstract}
 Based on random phase approximation (RPA), we numerically calculate dynamical structure factors of a balanced two-dimensional (2D) Fermi superfluid, and discuss their energy, momentum and interaction strength dependence in the 2D BEC-BCS crossover. At a small transferred momentum, a stable Higgs mode is observed in the unitary 2D Fermi superfluid gas where the particle-hole symmetry is not satisfied. Stronger interaction strength will make the visibility of the dispersion of Higgs mode harder to be observed. We also discuss the dimension effect and find that the signal of the Higgs mode in two dimension is more obvious than that in 3D case. At a large transferred momentum regime, stronger interaction strength will induce the weight of the molecules excitation increasing, while in verse the atomic one decreasing, which shows the pairing information of Fermi superfluid. The theoretical results qualitatively agree with the corresponding Quantum Monte Carlo data.
\end{abstract}

%
%
\submitto{\NJP}
%
%
%

\section{Introduction}
Two-dimensional (2D) Fermi atomic gas is a good platform to study many-body physics, where the strong correlation effect plays an important role in determining physical properties. Experimentally, the 2D Fermi gas has already been realized by freezing the motion of 3D Fermi gas in one chosen dimension \cite{Kirill10,Feld11,Bernd11,Cheng16}. The famous Feshbach resonances which had been achieved in 2D Fermi gas can also be used to tune the interaction of atoms \cite{Bernd11}, to investigate the 2D BEC-BCS crossover \cite{Cheng16,Boettcher16,Ries15,Sommer12,Dyke11}. Moreover, the physical properties of 2D Fermi gas in the whole BEC-BCS crossover region had obtained broad research attentions and interest \cite{Marsiglio15,Watanabe13,He15,Perali11,Makhalov14,Shi15,Galea16,Bertaina11,Anderson15,Lukas16,Matsumoto16,Mulkerin15,Astrakharchik04}, in which the dynamical excitation is a significant aspect to study the 2D Fermi superfluid gas.

Dynamical structure factors are the Fourier transformation of the density-density correlation function. They are interesting many-body physics quantities, and contain quite rich information about dynamics of the system, including collective excitations at a small transferred momentum and single-particle excitations at a large transferred momentum. Experimentally they can be directly obtained by the two-photon Bragg scattering spectroscopy according to measure the speed of center-of-mass of system. This technique had already been successfully used to study dynamical structure factors of 3D Fermi atomic gas \cite{Veeravalli08,Hoinka17,Hoinka12}. Many theoretical works had been done to understand dynamical structure factors of 3D Fermi gas \cite{Combescot06,Zou10,Zou16,Zou18,Hu18,Hu12,Kuhnle10,Watabe10}.

To date, there is no two-photon Bragg spectroscopy experiment on 2D superfluid Fermi gas, but in 2017, the Quantum Monte Carlo (QMC) had been used by E. Vitali and coworkers to study dynamical structure factors at the large transferred momentum of 2D superfluid Fermi gas by simulating the imaginary-time correlation function \cite{Vitali17,Vitali19}. However, they did not do calculation at the small transferred momentum, which is related to the interesting collective excitation modes. To now, there is no related work to study the collective excitation modes in 2D Fermi superfluid gas by the dynamical structure factors.

Generally spontaneous symmetry breaking will lead to the appearance of two typical collective excitation modes, a gapless Goldstone mode and a gapped Higgs mode. The Goldstone mode is related to the long-wavelength phase fluctuations of the order parameter, which appears when the continuous symmetries are broken. The Higgs mode is related to the amplitude modulation of order parameter, and its stability is ensured by the Lorentz invariance in high-energy physics. It had also been discovered in high-energy physics. Usually, the stable Higgs mode is a rare mode due to the presence of decay factors. In condensed matter physics, the Higgs mode had been proved by the Raman scattering experiments in BCS superconductors or superfluid with weak interaction where the particle-hole symmetry plays the same role with Lorentz invariance \cite{Sooryakumar80,Matsunaga13}. In ultracold atomic gases, the Higgs mode had been found in the strongly interacting Boson gas of optical lattice \cite{Bissbort11,Endres12}. In recent experiment of 3D fermionic gas $^{6}$Li, A. Behrle and coworkers found the evidence for the existence of stable Higgs mode in a strongly interacting Fermi superfluid gas where the particle-hole symmetry is not satisfied \cite{Behrle18}. By inducing a periodic modulation of the amplitude of the pair order parameter, they found an excitation resonance near twice the pair gap. Theoretically some preliminary evidence for the Higgs mode in 3D unitary Fermi superfluid gas had been found by calculating the spectral function of single particle \cite{Han16,Liu16,Pekker15}. They found that there are two sharp peaks on the spectral function of the amplitude fluctuation attributed to Goldstone and Higgs modes. Although they gave evidence of Higgs mode in the unitary region, the dispersion of Higgs mode in which is still an open question.

In this work, based on the RPA theory which provides a quantitative prediction in 3D Fermi gas \cite{Zou10,Zou16}, we study the dynamics of 2D Fermi superfluid gas by calculating dynamical structure factors from a small transferred momentum to large transferred momentum. Although the quantum fluctuation of 2D system is large, the RPA theory can still be expected to give a qualitatively reliable prediction. Our paper is organized as follows. In the sections 2 and 3, we introduce the Green's functions and the dynamical structure factors within RPA, respectively, then discuss the collective modes in section 4, and present the single particle excitations in section 5. The static structure factor is obtained in section 6. Moreover, we discuss the dimensional effect between 2D and 3D Fermi gas in section 7, and give a summary in section 8. Finally, in section 9, some calculated details are shown in appendix.
\section{Mean-field description with Green's functions}
For a balanced two-component Fermi superfluid  with a s-wave contact interaction, the Hamiltonian can be described by
\begin{eqnarray}\label{tUmodel}
H=\sum_{\sigma}\int d^{2}{\bf r}\Psi^{\dagger}_{\sigma}({\bf r})\left[-\frac{\nabla^{2}}{2m}-\mu\right]\Psi_{\sigma}({\bf r})
 +U\int d^{2}{\bf r}\Psi^{\dagger}_{\uparrow}({\bf r})\Psi^{\dagger}_{\downarrow}({\bf r})\Psi_{\downarrow}({\bf r})\Psi_{\uparrow}({\bf r}),\nonumber\\
\end{eqnarray}
where $\Psi_{\sigma}$ and $\Psi^{\dagger}_{\sigma}$ are annihilation and generation operators for spin-$\sigma$ component, respectively, $\mu$ is the chemical potential, and $U$ is the bare interatomic attractive interaction strength. Here and thereafter, we always set $\hbar=1$. In superfluid state, there are four different density operators, with two normal one $\hat{n}_{1}=\Psi^{\dagger}_{\uparrow}\Psi_{\uparrow}$ and $\hat{n}_{2}=\Psi^{\dagger}_{\downarrow}\Psi_{\downarrow}$, and anomalous pairing one and its complex conjugate $\hat{n}_{3}=\Psi_{\downarrow}\Psi_{\uparrow}$ and $\hat{n}_{4}=\Psi^{\dagger}_{\uparrow}\Psi^{\dagger}_{\downarrow}$, which is related to the pair order parameter by $\Delta=-U<\Psi_{\downarrow}\Psi_{\uparrow}>$.
Within the mean-field approximation, the four-operator term in the interaction Hamiltonian can be expressed as,
$\Psi^{\dagger}_{\uparrow}\Psi^{\dagger}_{\downarrow}\Psi_{\downarrow}\Psi_{\uparrow}=n_{1}\Psi^{\dagger}_{\downarrow}\Psi_{\downarrow}+n_{2}\Psi^{\dagger}_{\uparrow}\Psi_{\uparrow}+n_{3}\Psi^{\dagger}_{\uparrow}\Psi^{\dagger}_{\downarrow}+n_{4}\Psi_{\downarrow}\Psi_{\uparrow}$, and we use the mean value of operator $<\hat{n}_{\sigma}>$ to replace the operator $\hat{n}_{\sigma}$ itself. Therefore, the mean-field Hamiltonian in the momentum space reads
\begin{eqnarray}\label{momenspace}
H_{\rm MF}&=&\sum_{{\bf k},\sigma}\xi_{\bf k}\Psi^{\dagger}_{{\bf k}\sigma}\Psi_{{\bf k}\sigma}
-\sum_{{\bf k}}\left(\Delta^*\Psi_{{\bf k}\downarrow}\Psi_{-{\bf k}\uparrow}+\Delta\Psi^{\dagger}_{-{\bf k}\uparrow}\Psi^{\dagger}_{{\bf k}\downarrow}\right),
\end{eqnarray}
where the single particle spectrum is $\xi_{\bf k}=\epsilon_{\bf k}-\mu$, and $\epsilon_{\bf k}={\bf k}^{2}/(2m)$. Then we define the diagonal Green's function $G({\bf k},\tau-\tau^{'})=-\langle T \Psi_{{\bf k}\sigma}(\tau)\Psi^{\dagger}_{{\bf k}\sigma}(\tau^{'})\rangle$  and off-diagonal Green's function $\Gamma^{\dagger}({\bf k},\tau-\tau^{'})=-\langle T \Psi^{\dagger}_{-{\bf k}\uparrow}(\tau)\Psi^{\dagger}_{{\bf k}\downarrow}(\tau^{'})\rangle$, respectively. The diagonal and off-diagonal Green's functions can be expressed in BCS form as \cite{Zhao18},
 \numparts\begin{eqnarray}
G(\bf{k},\omega)&=&\frac{U^2_{\bf{k}}}{\omega-E_{\bf{k}}}+
\frac{V^2_{\bf{k}}}{\omega+E_{\bf{k}}},\\
\Gamma^{\dagger}(\bf{k},\omega)&=&
\frac{\Delta^*}{2E_{\bf{k}}}\left(\frac{1}{\omega-E_{\bf{k}}}-\frac{1}{\omega+E_{\bf{k}}}\right),
\end{eqnarray}\endnumparts
where $U^2_{\bf{k}}=\left[1+{\xi}_{\bf{k}}/E_{\bf{k}}\right]/2$, $V^2_{\bf{k}}=\left[1-{\xi}_{\bf{k}}/E_{\bf{k}}\right]/2$,
and the quasiparticle spectrum $E_{\bf{k}}=\sqrt{{\xi}^2_{\bf{k}}+{|\Delta|^{2}}}$. At zero temperature, chemical potential $\mu$ and order parameter $\Delta$ are calculated by particle-number equation
\begin{eqnarray}\label{nr}
 N=\sum_{\bf k}\left(1-\frac{\xi_{\bf k}}{E_{\bf k}}\right),
\end{eqnarray}
and order parameter equation
\begin{eqnarray}\label{gapr}
 \frac{1}{U}=-\sum_{\bf k}\frac{1}{2E_{\bf k}},
\end{eqnarray}
To eliminate the divergence of Eq. (\ref{gapr}) introduced by s-wave contact interaction, the bare interaction strength $U$ should be regularized by \cite{Liu2012}
\begin{eqnarray}\label{geff}
 \frac{1}{U}=-\sum_{\bf k}\frac{1}{2\epsilon_{\bf k}+E_{\rm b}},
\end{eqnarray}
where $E_{\rm b}$ is magnitude of binding energy, tuning which the BEC-BCS crossover in 2D Fermi gas can be realized. In general, the interaction strength of  Fermi gas can also be described by the s-wave scattering length. Therefore, the 2D scattering length $a_{\rm 2D}$  is related to the binding energy by $E_{\rm b}=4\hbar^{2}/(ma_{\rm 2D}^{2}e^{2\gamma})$, where $\gamma\simeq 0.577$ is the Euler's constant \cite{He15,Bertaina11}. It is important that $a_{\rm 2D}$ only changes from $0$ to $+\infty$ in the whole crossover, which is different from the 3D case. More details can be found in reference paper \cite{Galea16}. Here a parameter $\eta={\rm ln}(k_{\rm F}a_{\rm 2D})$, we call it inverse interaction strength. The BEC limit of tightly bound composite bosons corresponds to $\eta\ll1$, where $\eta\gg1$ corresponds to the BCS region of weak interaction where $a_{\rm 2D}$ is divergent. The strong-coupling regime between the BCS and BEC regimes is near $\eta=1$ (unitary region in 2D) \cite{He15,Shi15,Galea16,Bertaina11,Astrakharchik04,Vitali17}. In this paper, we just focus on several particular values of the interaction parameter  $\eta=0,0.5,1.0,1.5,1.96$, to do discussions from BEC to BCS region.

\section{Response functions and dynamical structure factors within RPA}
The RPA theory is a conventional method to calculate physical properties beyond the mean-field theory. Only in the frame of mean-field theory, it is not enough to give a correct prediction about the dynamical excitation of an interacting system, because it neglects the contribution from the fluctuation term of interaction Hamiltonion. The  RPA theory takes this fluctuation part back, and deal with it as a self-generated mean-field potential $\delta V^{\rm SC}$ experienced by particles \cite{Liu2004}
\begin{eqnarray}\label{deVsc}
\delta V^{\rm SC}=U\int d^{2}{\bf r}\left[\delta {\rm n_{4}}\hat{n}_3+\delta {\rm n_{3}}\hat{n}_4
 + \delta {\rm n}_{1}\hat{n}_2+\delta {\rm n}_{2}\hat{n}_1\right],
\end{eqnarray}
where  $\delta{\rm n}=[\delta{\rm n}_{1}, \ \delta{\rm n}_{2}, \ \delta{\rm n_{3}}, \ \delta{\rm n_{4}}]^{T}$ is the matrix of four particle density fluctuations.
Based on the linear response theory, when giving a weak external perturbation potential $V_{\rm ext}=[V_{1}, \ V_{2}, \ V_{3}, \ V_{4}]^{T}$ to the system, this density fluctuation $\delta n$ will be generated, it connects this external potential $V_{\rm ext}$ with
 \begin{eqnarray}\label{nxv}
\delta n=\chi V_{\rm ext},
\end{eqnarray}
 where $\chi$ is the response function matrix of the system, which is usually quite hard to be directively calculated. The RPA theory suggests us that we can define an effective potential  $V_{\rm eff} \equiv V_{\rm ext}+\delta V^{\rm SC}$. In the influence of $V_{\rm eff}$, the density fluctuation $\delta n$ is connected to this effective potential $V_{\rm eff}$ by
\begin{eqnarray}\label{veff}
\delta n=\chi^{0} V_{\rm eff},
\end{eqnarray}
where $\chi^0$ is the response function matrix in the mean-field theory, whose calculation is very easy. By this treatment introduced by RPA theory,
the response function $\chi$ can be obtained by its connection to the mean-field response function $\chi^0$
 \begin{eqnarray}\label{chi}
 \chi(q,i\omega_{n})=\frac{\chi^{0}(q,i\omega_{n})}{\hat{1}-\chi^{0}(q,i\omega_{n})UG},
\end{eqnarray}
where $G=\sigma_{0}\otimes\sigma_{x}$ is a direct product of two Pauli matrices $\sigma_{0}$ and $\sigma_{x}$, $\sigma_{0}$ is the unit matrix and the unit matrix $\hat{1}=\sigma_{0}\otimes\sigma_{0}$.

The matrix expression of mean-field response function $\chi^{0}(q,i\omega_{n})$ reads
\begin{eqnarray}\label{matrix}
\chi^{0}(q,i\omega_{n})=\left[
\begin{array}{cccccc}
&\chi^{0}_{11}&\chi^{0}_{12}&\chi^{0}_{13}&\chi^{0}_{14}\\
&\chi^{0}_{21} &\chi^{0}_{22}&\chi^{0}_{23}&\chi^{0}_{24}\\
&\chi^{0}_{31}&\chi^{0}_{32}&\chi^{0}_{33}&\chi^{0}_{34}\\
&\chi^{0}_{41}&\chi^{0}_{42}&\chi^{0}_{43}&\chi^{0}_{44}\\
\end{array}
\right].
 \end{eqnarray}
 These 16 matrix elements are determined by the corresponding density-density correlation functions which can be obtained by defining corresponding Green's functions. In fact, as a result of the symmetry of system, only 6 matrix elements are independent, i.e., $\chi^{0}_{11}=\chi^{0}_{22}$, $\chi^{0}_{12}=\chi^{0}_{21}=-\chi^{0}_{33}=-\chi^{0}_{44}$,
 $\chi^{0}_{31}=\chi^{0}_{32}=\chi^{0}_{14}=\chi^{0}_{24}$,
 $\chi^{0}_{41}=\chi^{0}_{42}=\chi^{0}_{13}=\chi^{0}_{23}$. These elements have been obtained in the appendix part of this paper.
In particular, $\chi^{0}_{43}$ and $\chi^{0}_{34}$ are divergent when $k\to \infty$, and the denominator of Eq. (\ref{chi}) is just a proper way to eliminate their divergences. Since we can get two convergent response functions $\widetilde{\chi}_{43}=\chi^{0}_{43}-1/U$ and $\widetilde{\chi}_{34}=\chi^{0}_{34}-1/U$, we find the density response function is expressed as
\begin{eqnarray}\label{xdw}
\chi_{D}=2\left(\chi^{0}_{11}+\chi^{0}_{12}\right)
-4\frac{{\left(\chi^{0}_{31}\right)}^{2}\widetilde{\chi}_{43}+{\left(\chi^{0}_{41}\right)}^{2}\widetilde{\chi}_{34}
+2\chi^{0}_{12}\chi^{0}_{31}\chi^{0}_{41}}{\widetilde{\chi}_{34}\widetilde{\chi}_{43}-{\left(\chi^{0}_{12}\right)}^{2}}.
\end{eqnarray}
According to the fluctuation-dissipation theory, the density dynamical structure factor $S(q,{\omega})$ is connected to the imaginary part of the density response function $\chi_{D}$ by
\begin{eqnarray}\label{sqw}
  S(q,{\omega})&=&-\frac{1}{\pi}{\rm Im}\chi_{D}(q,i\omega_{n}\to \omega+i\delta),
 \end{eqnarray}
where $q$ and $\omega$ are the transferred momentum and energy, respectively. $\delta$ is a small positive number (usually we set $\delta=0.001$). $S(q,\omega)$ satisfies the famous f-sum rule $\int d\omega \omega S(q,{\omega})=Nq^{2}/(2m)$.
And the spin dynamical structure factor $S_{S}(q,{\omega})$ reflects the excitations related to spin, and similarly it is connected to the imaginary part of spin response function
 \begin{eqnarray}\label{xdw}
  \chi_{S}(q,i\omega_{n})&=&2(\chi^{0}_{11}-\chi^{0}_{12}).
 \end{eqnarray}
with
\begin{eqnarray}\label{sqw}
  S_S(q,{\omega})&=&-\frac{1}{\pi}{\rm Im}\chi_{S}(q,i\omega_{n}\to \omega+i\delta).
 \end{eqnarray}
 \section{Phonon and Higgs modes}
Now we discuss dynamical structure factors at a small transferred momentum region to obtain the information of collective excitations. In a uniform system with  density $n$, we can use Fermi wave vector $k_{\rm F}=\sqrt{2\pi n}$ and Fermi energy $E_{\rm F}={k_{F}}^{2}/(2m)$ as units of momentum and energy.

\begin{figure}[ht]
  \centering
  \includegraphics[width=0.8\textwidth]{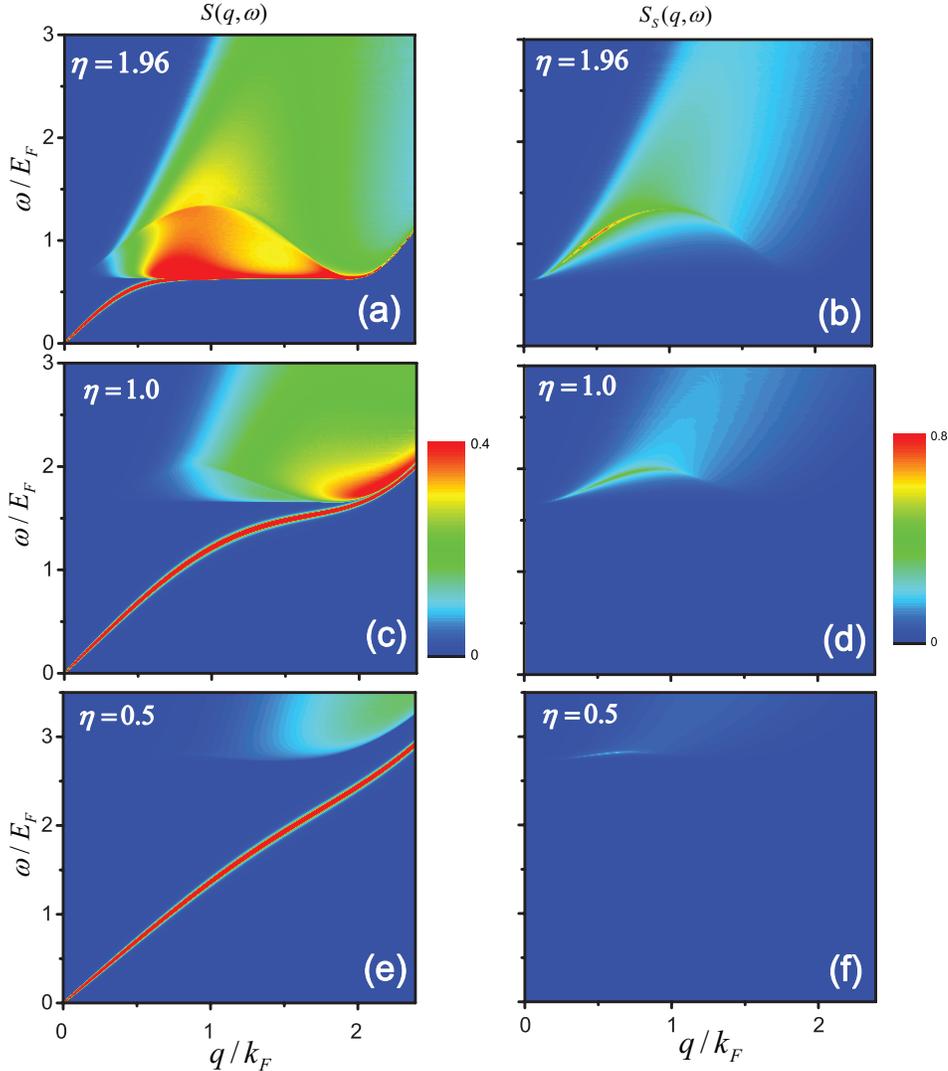}
  \vspace{-3mm}
  \caption{The color maps of density $S(q,{\omega})$ (left) and spin $S_{S}(q,{\omega})$ (right) dynamical structure factors from $q=0$ to $q=2.4k_{F}$ for interaction strength $\eta=1.96$ (top), $\eta=1.0$ (middle) and $\eta=0.5$ (bottom), respectively.}
  \vspace{-3mm}
  \label{fig1} 
\end{figure}

We calculate the energy and momentum dependence of density dynamical structure factor $S(q,{\omega})$ and spin dynamical structure factor $S_{S}(q,{\omega})$ in the 2D BEC-BCS crossover. We choose three typical inverse interaction strength parameters, $\eta=1.96$ (BCS), $\eta=1.0$ (unitarity) and $\eta=0.5$ (BEC), and plot $S(q,{\omega})$ (left) and $S_{S}(q,{\omega})$ (right) in Fig. \ref{fig1}.
The dispersions of $S(q,{\omega})$ and $S_{S}(q,{\omega})$ have a dramatic variation in the whole crossover. First, the $S(q,{\omega})$ has a sharp narrow peak at the small transferred momentum region, and its dispersion starts from zero energy, then increases almost linearly. This linear dispersion is related to the collective phonon mode, and the slope of phonon dispersion at $\omega \rightarrow 0$ is the sound speed $c_{\rm s}$, $c_{\rm s}=\omega/q$. For $\eta=1.96$, $c_{\rm s}\approx0.714v_{\rm F}$, which is close to the analytical result of free Fermi gas $1/{\sqrt{2}}v_{\rm F}$. Second, at the transferred energy $\omega=2\Delta$, a horizontal threshold appears, indicates the minimum energy to break a Cooper pair, whose values are $2\Delta\approx0.63, 1.65, 2.73E_F$ for $\eta=1.96, 1.0, 0.5$, respectively. Therefore, by virtue of the measurement of density dynamical structure factor, the magnitude of pair gap can be obtained. Third, above the horizontal threshold energy, another collective excitation of $S(q,{\omega})$ appears, which starts from twice the pair gap at $q=0$, behaves almost quadratically with the transferred momentum $q$, and then disappears at around $q=2.0k_{F}$. The mode is related to Higgs mode.

\begin{figure}[ht]
  \centering
  \includegraphics[width=0.7\textwidth]{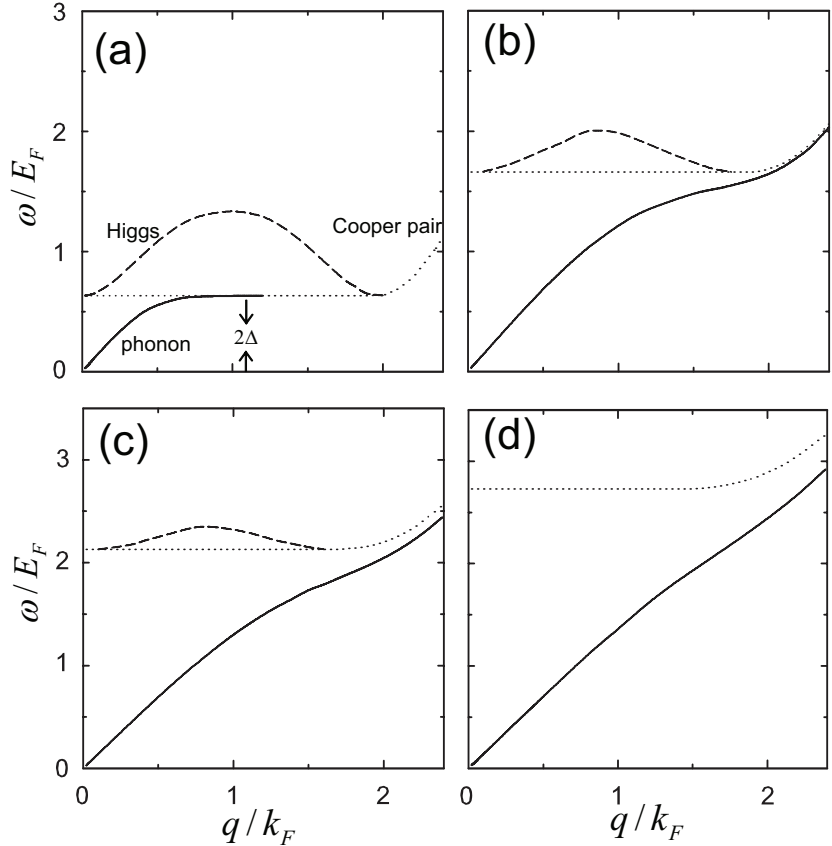}
  \vspace{-3mm}
  \caption{The dispersion of $S(q,{\omega})$  from $q=0$ to $q=2.4k_{F}$ for interaction strength (a) $\eta=1.96$, (b) $\eta=1.0$, (c) $\eta=0.75$ and (d) $\eta=0.5$, respectively. The solid line is the phonon, the dashed line corresponds to the Higgs mode while the dotted line is the Cooper pair excitations.}
  \vspace{-3mm}
  \label{fig2} 
\end{figure}

The Higgs mode is obvious in the BCS region, where its dispersion at the BCS limit can be expressed as $\omega^{2}=v_{F}^{2}q^{2}/2+(2\Delta)^{2}$ at the small transferred momentum region \cite{Han16}. However, in the unitary region, the Higgs mode is flattened and suppressed, which shows that the Higgs mode is strongly influenced by interaction strength. Stronger interaction strength will make the visibility of the dispersion of Higgs mode harder to be observed. The Higgs mode can also be observed from the spin dynamical structure factor $S_{S}(q,{\omega})$, which is the same as the phenomenon in $S(q,{\omega})$. In Fig. \ref{fig2}, we plot dispersions of phonon and Higgs mode to summarize our main results of the $S(q,{\omega})$ of 2D Fermi superfluid gas from $q=0$ to $q=2.4k_{F}$ for (a) $\eta=1.96$, (b) $\eta=1.0$, (c) $\eta=0.75$ and (d) $\eta=0.5$, respectively.
It is thus shown that the collective excitations in 2D Fermi superfluid gas are sensitive on the interaction strength, especially for the Higgs mode. In the unitary region, the intensity of the Higgs mode is obviously suppressed and the dispersion is flattened, which is consistent with the experimental result of the 3D Fermi superfluid gas \cite{Behrle18}.  Go on increasing of interaction strength, the signal of Higgs mode is becoming weaker. Moreover, in the deep BEC region, the Higgs mode disappears on the density dynamical structure factor, in consistent with the results of other groups \cite{Han16,Liu16,Pekker15}.

\begin{figure}[ht]
  \centering
  \includegraphics[width=0.8\textwidth]{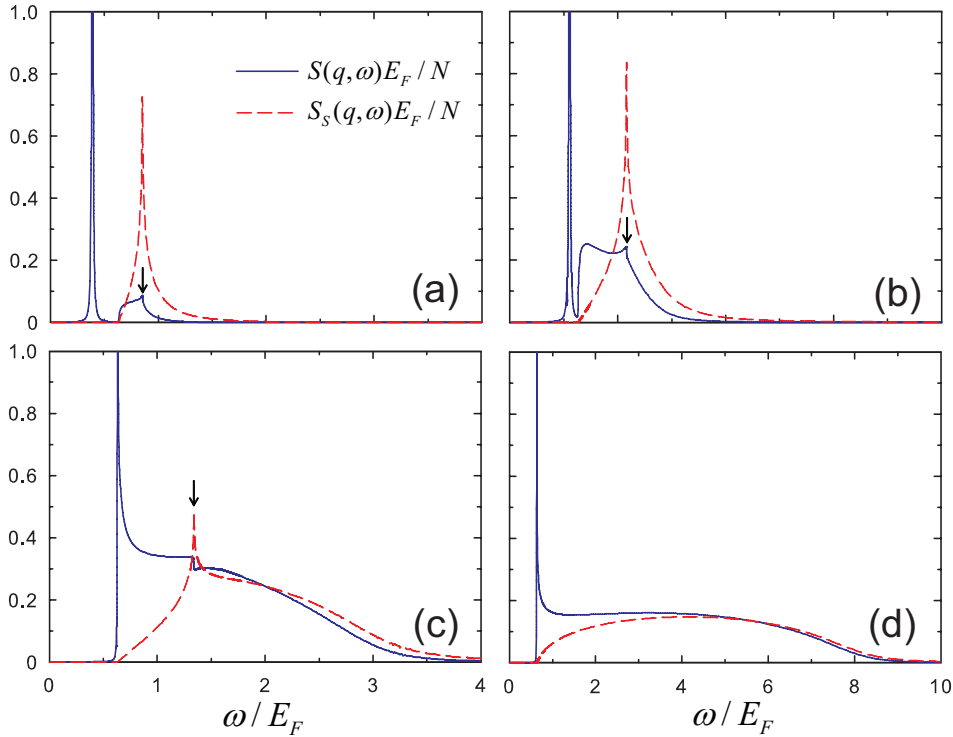}
  \vspace{-3mm}
  \caption{The dynamical structure factors $S(q,{\omega})$ (blue solid line) and $S_{S}(q,{\omega})$ (red dashed line) as a function of ${\omega}$ for (a) $q=0.3k_{F}$, (b) $q=0.5k_{F}$, (c) $q=1.0k_{F}$ and (d) $q=2.0k_{F}$ with parameters $\eta=1.96$.}
  \vspace{-3mm}
  \label{fig3} 
\end{figure}

 To show clearly the energy and momentum dependence of dynamical structure factors in the collective excitation regime, we have calculated the energy dependence of $S(q,{\omega})$ and $S_{S}(q,{\omega})$ at some fixed transferred momentums. Related results of $S(q,{\omega})$ (blue solid line) and $S_{S}(q,{\omega})$ (red dashed line) as a function of ${\omega}$ for (a) $q=0.3k_{F}$, (b) $q=0.5k_{F}$, (c) $q=1.0k_{F}$ and (d) $q=2.0k_{F}$ in BCS region ( $\eta=1.96$) are plotted in Fig. \ref{fig3}.
From left to right, the phonon mode, Cooper-pair excitation and Higgs mode are displayed in order. In particular, above the horizontal threshold energy ($2\Delta$), a characteristic structure (a peak or a jump ) of $S(q,{\omega})$ marked by the arrows locates the Higgs mode, which is a peak signal in $S_{S}(q,{\omega})$ at $q=0.3k_{F}$ and $q=0.5k_{F}$. Another broad peak arises at $q=1.0k_{F}$ which corresponds to the particle-hole excitations while the sharp peak of Higgs mode is strongly suppressed. When at $q=2.0k_{F}$, the peak of Higgs mode disappears, leaving the broad peak of particle-hole excitations.

\begin{figure}[ht]
  \centering
  \includegraphics[width=0.6\textwidth]{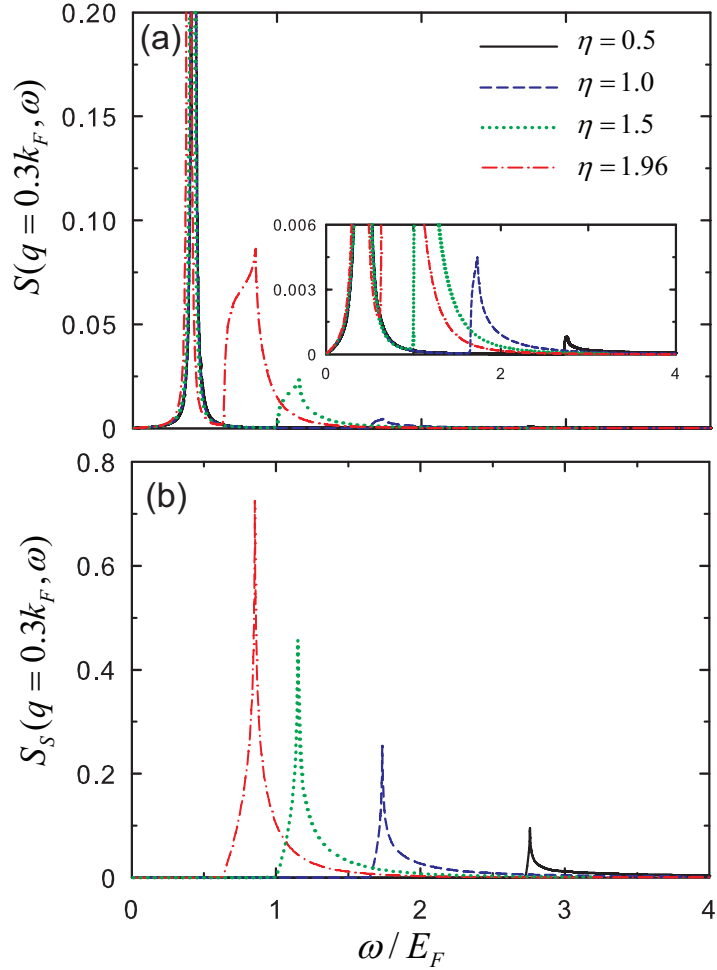}
  \vspace{-3mm}
  \caption{(a) $S(q=0.3k_{F},{\omega})$ and (b) $S_{S}(q=0.3k_{F},{\omega})$ as a function of $\eta$, respectively. The inset in
the upper panel highlights the density response at the low-energy region.}
  \vspace{-3mm}
  \label{fig4} 
\end{figure}

Furthermore, the shape of dynamical structure factors is strongly reconstructed by the interaction strength. We plot $S(q,{\omega})$ and $S_{S}(q,{\omega})$ as a function of interaction strength $\eta$ at $q=0.3k_{F}$ in Fig. \ref{fig4}a and Fig. \ref{fig4}b, respectively.
Decreasing $\eta$ (interaction increases), the low-energy phonon part of the $S(q=0.3k_{F},{\omega})$ changes little, but the higher energy part has an obvious and continuous variation, i.e., the pair gap increases which leads to that both the horizontal threshold part and Higgs mode move to the larger energy region and their intensities are strongly suppressed, compared with phonon excitation. The signal of Higgs mode in $S(q=0.3k_{F},{\omega})$ disappears in the BEC region.  Moreover, the peak signal of Higgs mode in $S_{S}(q=0.3k_{F},{\omega})$ decreases its strength, and the corresponding peak position moves to the larger energy region.

\section{Single particle excitations}
\begin{figure}[ht]
  \centering
  \includegraphics[width=0.8\textwidth]{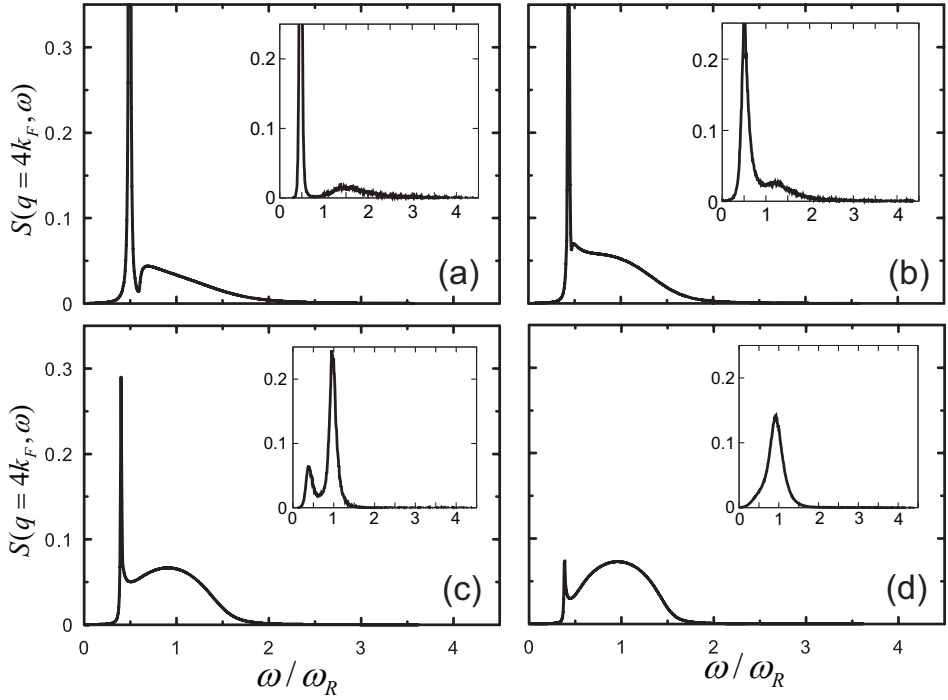}
  \vspace{-3mm}
  \caption{Density dynamical factor $S(q,{\omega})$ as a function of $\omega$ for (a) $\eta=0$, (b)$\eta=0.5$, (c) $\eta=1.0$, and (d) $\eta=1.5$. Inset: the corresponding QMC data \cite{Vitali17}.}
  \vspace{-3mm}
  \label{fig5} 
\end{figure}

\begin{figure}[ht]
  \centering
  \includegraphics[width=0.8\textwidth]{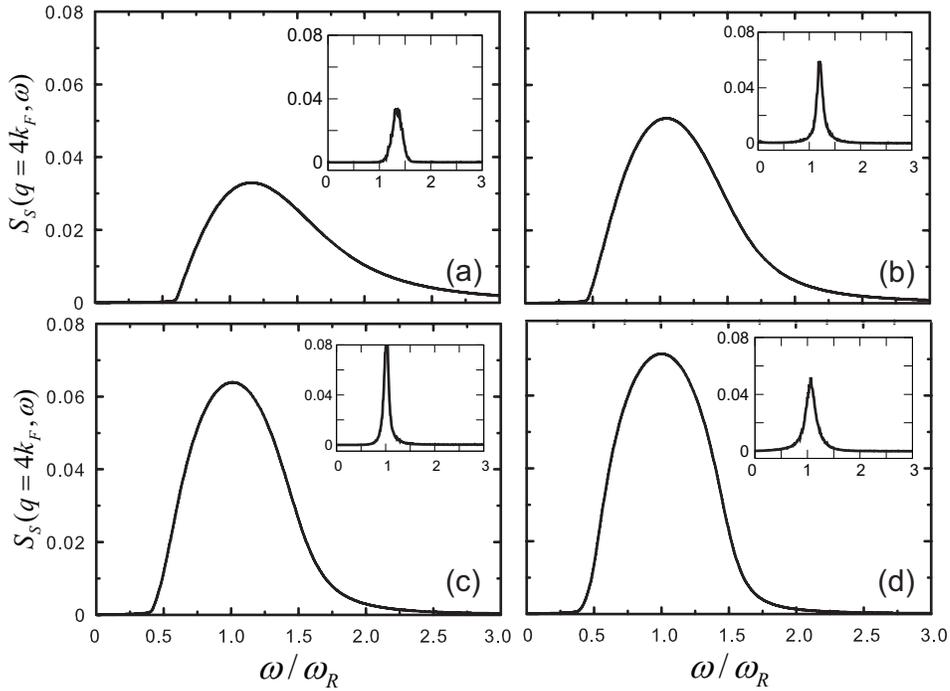}
  \vspace{-3mm}
  \caption{Spin dynamical factor $S_{S}(q,{\omega})$ as a function of $\omega$ for (a) $\eta=0$ , (b) $\eta=0.5$, (c) $\eta=1.0$, and (d) $\eta=1.5$. Inset: the corresponding QMC data \cite{Vitali17}.}
  \vspace{-3mm}
  \label{fig6} 
\end{figure}

In large transferred momentum region, dynamical structure factors provide information of Cooper pair molecules and atoms excitations, especially for the Cooper pair breaking excitation. At $q=4k_{F}$, we plot the energy dependence of $S(q,{\omega})$ and $S_{S}(q,{\omega})$ for (a) $\eta=0$, (b) $\eta=0.5$, (c) $\eta=1.0$, and (d) $\eta=1.5$ in Fig. \ref{fig5} and Fig. \ref{fig6}, respectively, and compare with the corresponding QMC results (inset) \cite{Vitali17}. A characteristic energy $\omega_{R}={{\bf q}^{2}/(2m)}=16E_{F}$ is used as the unit of transferred energy during comparison. We use the same small quantity $\delta=0.1$ in QMC simulations.

The density dynamical structure factor $S(q=4k_{F},{\omega})$ is composed of a sharp molecule peak around $\omega\approx \omega_{R}/2$ and an atomic peak around $\omega\approx \omega_{R}$. At BEC region ($\eta=0$), the weight of the molecules peak is high while the atomic peak around $\omega_{R}$ is strongly suppressed and disappear in the deep BEC region. Increasing the inverse interaction strength $\eta$, the weight of the molecules peak decreases but the atomic peak appears and increases quickly. The 2D BEC-BCS crossover can be well understood through intensity change of atomic and molecule peaks of $S(q,{\omega})$.
The spin dynamical structure factor $S_{S}(q,{\omega})$ reaches its maximum around the energy $\omega_{R}$. With the increase of $\eta$ (decreasing interaction strength), the peak position of $S_{S}(q,{\omega})$ moves to a lower energy region. These theoretical results are similar to the 3D case, and qualitative with the corresponding QMC data \cite{Vitali17}.

\section{Static structure factor}
\begin{figure}[ht]
  \centering
  \includegraphics[width=0.5\textwidth]{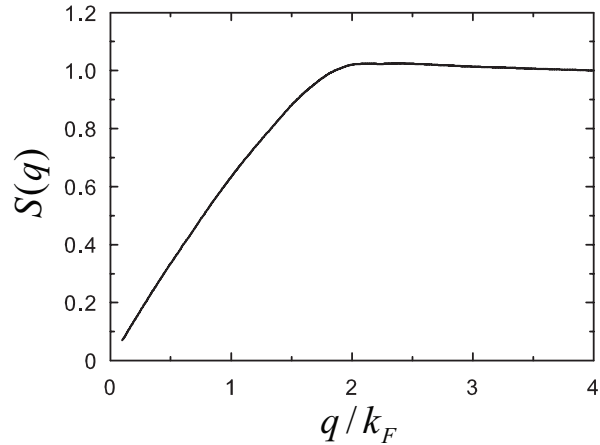}
  \vspace{-3mm}
  \caption{The static structure factor $S(q)$ as a function of transferred momentum with $\eta=1.96$.}
  \vspace{-3mm}
  \label{fig7} 
\end{figure}

The static structure factor $S(q)$ can be obtained by integrating the transferred energy over the dynamical structure factor $S(q,{\omega})$,
\begin{eqnarray}\label{static}
 S(q)=\int d\omega S(q,\omega).
\end{eqnarray}
The static structure factor $S(q)$ is shown in Fig. \ref{fig7} with $\eta=1.96$.
We find the static structure factor $S(q)$ increases linearly at the small transferred momentum region, and reaches the maximum at around $q=2.0k_{F}$, then decreases to constant at the large transferred momentum, $S(q\to \infty)=1$ which is a model-independent quantity.

\section{Dimensional effect between 2D and 3D Fermi gas}
\begin{figure}[ht]
  \centering
  \includegraphics[width=0.8\textwidth]{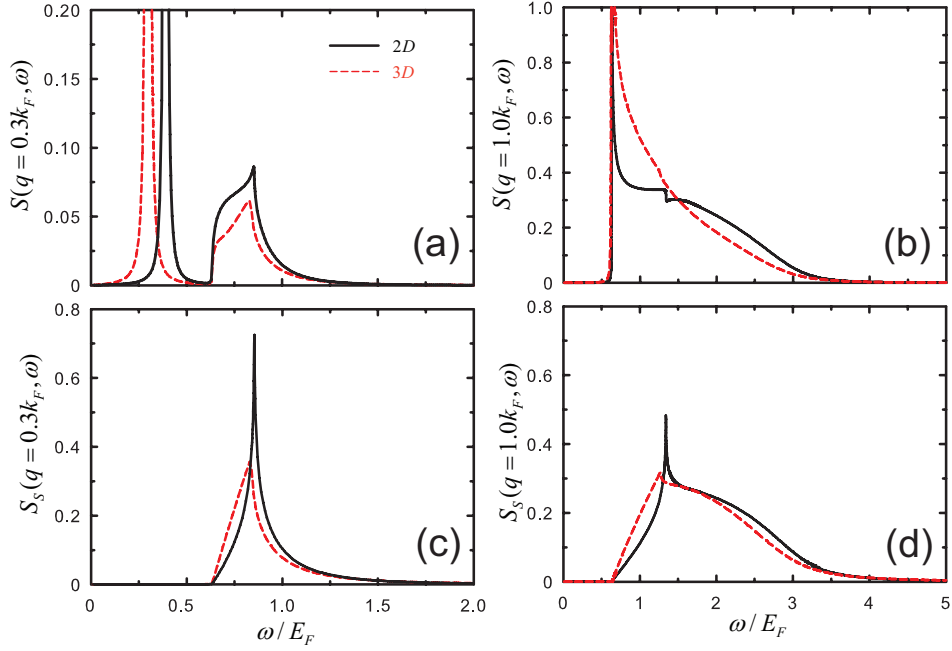}
  \vspace{-3mm}
  \caption{The $S(q,{\omega})$ (top) and $S_{S}(q,{\omega})$ (bottom) in 2D and 3D Fermi superfluid gas, respectively, for $q=0.3{k_{F}}$ (left) and $q=1.0{k_{F}}$ (right).}
  \vspace{-3mm}
  \label{fig8} 
\end{figure}

Now we discuss the influence of spatial dimension by calculating the dynamical structure factors for both 2D and 3D case. In order to facilitate the comparison of different spatial dimension, approximately we choose the same pair gap intensity to stand for almost the same interaction strength in two cases. For 2D Fermi superfluid gas, $\eta={\rm ln}(k_{F}a_{2D})=1.96$ while $1/{k_{F}a_{3D}}=-0.692$ in 3D case. In Fig. \ref{fig8}, we have calculated the $S(q,{\omega})$ (top) and $S_{S}(q,{\omega})$ (bottom) in 2D and 3D Fermi superfluid gas for $q=0.3{k_{F}}$ (left) and $q=1.0{k_{F}}$ (right), respectively.
Generally the behavior of two different spatial dimensions is almost similar to each other. However, the Higgs mode is suppressed in 3D case, and 2D dimension can do help to observe an obvious Higgs excitation, even at a relatively large transferred momentum $q=k_F$, where almost no signal in 3D case.  In Fig. 9, the contourplot of $S(q,{\omega})$  and  $S_{S}(q,{\omega})$ of 3D Fermi gas at $1/{k_{F}a_{3D}}=-0.692$  are also been finished.
\begin{figure}[ht]
  \centering
  \includegraphics[width=0.8\textwidth]{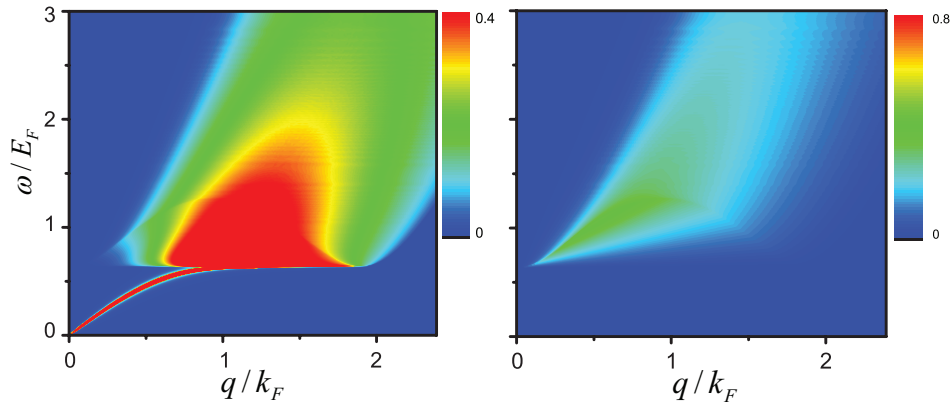}
  \vspace{-3mm}
  \caption{The color maps of $S(q,{\omega})$ (left) and  $S_{S}(q,{\omega})$ (right) from $q=0$ to $q=2.4k_{F}$ for interaction strength $1/{k_{F}a_{3D}}=-0.692$, respectively.}
  \vspace{-3mm}
  \label{fig9} 
\end{figure}
 It is shown that there is a quite weak Higgs excitation in 3D Fermi gas by comparing with 2D result. Therefore, the lower dimension is more conducive to the study of Higgs mode. Moreover, the slope of the linear phonon dispersion in 3D Fermi gas is smaller than the 2D case, which indicates the larger sound speed in 2D case, in consistent with free Fermi gas results.

\section{Summary }
In conclusion, the dynamical structure factors of 2D Fermi superfluid gas are studied in the whole 2D BEC-BCS crossover based on the RPA theory. These theoretical results predict the collective excitations at a small transferred momentum, the collective phonon and Higgs mode are observed clearly. In particular, the Higgs mode indeed exists in the unitary region where the particle-hole symmetry is destroyed. Moreover, the Higgs mode is flattened and suppressed as the interaction strength increases. Stronger interaction strength will make the visibility of the dispersion of Higgs mode harder to be observed. And the signal of the Higgs mode is more obvious in 2D Fermi superfluid gas than 3D case.
 At a large transferred momentum, there is a strong enhancement of weight transfer in density dynamic structure factor from the atomic excitation to molecular one when increasing the interaction strength. These results of the large transferred momentum region are in qualitative agreement with the corresponding QMC data.
\section{Appendix}
 Based on the mean-field theory by calculating the corresponding Green's functions in 2D Fermi gas, we can obtain the correlation functions $\chi^{0}_{11},\chi^{0}_{12},\chi^{0}_{31},\chi^{0}_{41},\chi^{0}_{43},\chi^{0}_{34}$ as,
\begin{eqnarray}\label{a1}
 \chi^{0}_{11}=\frac{1}{4}\sum_{k} \left|V_{kq}\right|^2F_{1}(k,q)\nonumber
\end{eqnarray}
\begin{eqnarray}\label{b}
\chi^{0}_{12}=\sum_{k} \frac{|\Delta|^{2}}{4E_{k}E_{k+q}}F_{1}(k,q) \nonumber
\end{eqnarray}
\begin{eqnarray}\label{c}
\chi^{0}_{31}&=&\frac{1}{8} \sum_{k} \frac{\Delta}{E_{k}E_{k+q}}\left [\left(\xi_{k}+\xi_{k+q}\right)F_{1}(k,q)
 +\left(E_{k+q}+E_{k}\right)F_{2}(k,q)\right]\nonumber
\end{eqnarray}
\begin{eqnarray}\label{ct}
\chi^{0}_{41}&=&\frac{1}{8} \sum_{k} \frac{\Delta}{E_{k}E_{k+q}}\left [\left(\xi_{k}+\xi_{k+q}\right)F_{1}(k,q)
 -\left(E_{k+q}+E_{k}\right)F_{2}(k,q)\right]\nonumber
\end{eqnarray}
\begin{eqnarray}\label{h}
\chi^{0}_{43}&=&\frac{1}{4}\sum_{k}\left[\left|U_{kq}\right|^{2}F_{1}(k,q)-\left(\frac{\xi_{k}}{E_{k}}+\frac{\xi_{k+q}}{E_{k+q}}\right)F_{2}(k,q)\right]\nonumber
\end{eqnarray}
\begin{eqnarray}\label{ht}
\chi^{0}_{34}&=&\frac{1}{4}\sum_{k}\left[\left|U_{kq}\right|^{2}F_{1}(k,q)+\left(\frac{\xi_{k}}{E_{k}}-\frac{\xi_{k+q}}{E_{k+q}}\right)F_{2}(k,q)\right]\nonumber
.
\end{eqnarray}
where $|U_{kq}|^2=1+\xi_{k}\xi_{k+q}/(E_{k}E_{k+q})$ and $|V_{kq}|^2=1-\xi_{k}\xi_{k+q}/(E_{k}E_{k+q})$, the corresponding functions  $F_{1}(k,q)$,  $F_{2}(k,q)$ are shown as
\begin{eqnarray}\label{Fkq}
 F_{1}(k,q)&=&\frac{1}{i\omega_{n}-(E_{k}+E_{k+q})}-\frac{1}{i\omega_{n}+(E_{k}+E_{k+q})}\nonumber\\
 F_{2}(k,q)&=&\frac{1}{i\omega_{n}-(E_{k}+E_{k+q})}+\frac{1}{i\omega_{n}+(E_{k}+E_{k+q})}\nonumber\\
 .
\end{eqnarray}
\section*{Acknowledgments}
The authors would like to thank Prof. Shiping Feng and Supeng Kou for helpful discussions.
This work was supported by the funds from the National
Natural Science Foundation of China under Grant Nos.11547034,11804177 and the Shandong Provincial Natural Science Foundation, China, Grant No. ZR2018BA032.

\end{document}